\documentclass[conference]{IEEEtran}

\makeatletter
\def\endthebibliography{%
	\def\@noitemerr{\@latex@warning{Empty `thebibliography' environment}}%
	\endlist
}
\makeatother

\usepackage{cite}
\usepackage{amsmath,amssymb,amsfonts}
\usepackage{algorithmic}
\usepackage{graphicx}
\usepackage{caption}
\usepackage{textcomp}
\usepackage{xcolor}
\usepackage{float}
\usepackage{varwidth}
\usepackage{subfloat}
\usepackage{placeins}
\usepackage{tikz}
\usepackage{pgfplots}

\usetikzlibrary{arrows,backgrounds}
\usepgflibrary{shapes.multipart}
\begin{document}

\title{Deep Autoencoders for DOA Estimation of Coherent Sources using Imperfect Antenna Array \\
}
\author{\IEEEauthorblockN{Aya Mostafa Ahmed, Omar Eissa and
		Aydin Sezgin}
	\IEEEauthorblockA{Institute of Digital Communication Systems \\Faculty of Electrical and Computer Engineering\\ Ruhr-Universität Bochum, Germany\\
		Email: \{aya.mostafaibrahimahmad; omar.eissa; aydin.sezgin\}@rub.de}}

 \maketitle

\begin{abstract}
In this paper a robust algorithm for DOA estimation of coherent sources in presence of antenna array imperfections is presented. We exploit the current advances of deep learning to overcome two of the most common problems facing the state of the art DOA algorithms (i.e. coherent sources and array imperfections). We propose a deep auto encoder (AE) that is able to correctly resolve coherent sources without the need of spatial smoothing, hence avoiding possible processing overhead and delays. Moreover, we assumed the presence of array imperfections in the received signal model such as mutual coupling, gain/ phase mismatches, and position errors. The deep AE is trained using the covariance matrix of the received signal, where it alleviates the effect of imperfections, and at the same time act as a filters for the coherent sources. The results show significant improvement compared to the methods used in the literature.
\end{abstract}


\section{Introduction}
Direction-of-arrival (DOA) estimation is a common problem in different fields, including wireless communication, astronomical observation, radar and sonar. There are various challenges facing DOA estimation such as accuracy and precision of estimates in non-ideal scenarios, e.g., multipaths and antennas imperfections. There are many DOA estimation techniques that can be classified to multiple categories. On one hand, there are the conventional methods which depends on the locations of peaks in the spatial spectrum to determine the DOA, i.e., delay and sum beamforming and minimum variance distortionless response (MVDR) \cite{mvdr}. On the other hand there are also the subspace methods  e.g., MUltiple SIgnal Classification (MUSIC) \cite{b7} which depends on the eigenstructure of the spatial correlation matrix, offering high resolution DOA estimation. There also exists parametric-based approach like the maximum likelihood (ML) technique, which uses a statistical structure for the process of generating data \cite{b3}. However, the application of those methods in real life applications is very limited, as it requires the accurate knowledge of the received signal without any errors. Hence, they are limited to specific scenarios where the antenna array measurements are ideal and the sources are not correlated. For example, the MUSIC algorithm has the ability to detect and measure multiple sources at the same time with high precision measurement only under ideal array conditions in a non multipath environment.  However its performance degrades significantly in the existence of array imperfections or when the sources are coherent (fully correlated) due to multi-path. In such scenario, the rank of the signal covariance matrix is less than the number of impinging signals, which significantly affects the estimation performance. To solve this problem, spatial smoothing techniques have been widely used to remove the coherence between sources or de-correlate their signals by reconstructing a full rank covariance matrix before going through the estimation algorithm \cite{SSMUSIC}. However, this adds processing overhead, making it difficult to use in real time applications. Moreover, spatial smoothing only solves the coherence problem, and does not solve other problems like antenna array imperfections which is a common problem in practice. As a matter of fact, those imperfections generally occur in practical applications due to the non-idealities in the antenna array such as mutual coupling, gain / phase gradual changes over time and as well as changes in the antenna locations. For instance,  mutual coupling occurs due to interference from nearby antennas during transmitting \cite{fried}, while gain and phase inconsistency can result from the aging of electrical components or thermal effects. All of those factors change the antenna array response, causing significant degradation in the performance of DOA algorithms like MUSIC \cite{fried}. 
 Many approaches in the literature addressed those problems using auto calibration algorithms as in \cite{MCsol1,MCsol2}. However, such algorithms require the prior knowledge of the imperfections formulations, which can be challenging in practical applications. Other approaches used the latest advances in machine learning, and deep learning to solve this problem as in \cite{MLpaper}. However, the authors in \cite{MLpaper} only focused on the imperfections problem, where they proposed a neural network followed by a classifier. However, they didn't take into consideration potential correlation among the sources that can result from multipath. 
 \\
In this paper we propose a deep neural network framework for DOA estimation that is generic and robust against both array imperfections, along with the existence of coherent sources. The results show that our DNN can correctly estimate the directions of spatially close coherent sources without any prior spatial smoothing techniques. Furthermore, our algorithm showed significantly better results compared to the state of the art methods for coherence removal as  spatial smoothing MUSIC algorithm \cite{SSMUSIC}, and to similar algorithms for imperfections as aforementioned approach found in \cite{MLpaper}.  
\section{System Model}
\subsection{Array Imperfections}
Assume that $K$ sources are impinging on a receive array consisting of $M$  antennas, whose DOA are $\theta_1,\ldots,\theta_K$.  The received signal is sampled at $N$ uniquely spaced time instants $t_1,\ldots,t_N$ to obtain multiple snapshots gathered in matrix $\mathbf{Z} = [\mathbf{z}(t_1),\ldots,\mathbf{z}(t_N)]$, with
\begin{equation}
\boldsymbol{z}\left(t_{n}\right)=\sum_{k=1}^{K} \boldsymbol{a}\left(\theta_{k}\right) x_{k}\left(t_{n}\right)+\boldsymbol{w}\left(t_{n}\right), \quad \text { for } n=1, \ldots, N .
\label{received_signal}
\end{equation} 
$x_k(t_n)$ is the transmit waveform of the $k$ th source, and $\boldsymbol{w}\left(t_{n}\right)$ is the zero-mean Gaussian noise. $\boldsymbol{a}\left(\theta_{k}\right)$ denotes the imperfection free steering vector and is defined as
\begin{equation}
\boldsymbol{a}\left(\theta_{k}\right)=\frac{1}{\sqrt{M}}\left[1, e^{-j 2 \pi \frac{d}{\lambda} \sin \theta_{k}}, \cdots, e^{-j 2 \pi \frac{d}{\lambda}\left(M-1\right) \sin \theta_{k}}\right]^{T}
\end{equation}
The covariance matrix of the received signal $\boldsymbol{z}$ is
\begin{equation}
\label{covmatrix}
{\boldsymbol{R}}=\frac{1}{N} \sum_{n=1}^{N} \boldsymbol{z}\left(t_{n}\right) \boldsymbol{z}^{H}\left(t_{n}\right).
\end{equation}
The model in \eqref{received_signal} is the idealistic received signal without any imperfections in the antenna array, which is commonly used in the literature. However, it is quite impractical. Hence, we re-define \eqref{received_signal} as  
\begin{equation}
\label{errRx}
\boldsymbol{z}\left(t_{n}\right)=\sum_{k=1}^{K} \boldsymbol{a}\left(\theta_{k}, \boldsymbol{e}\right) x_{k}\left(t_{n}\right)+\boldsymbol{w}\left(t_{n}\right), \quad n=1, \ldots, N
\end{equation}
 where $\boldsymbol{a}\left(\theta_{k}, \boldsymbol{e}\right)$ is the array response after adding the array imperfections. Here we consider gain and phase errors ($\mathbf{e}_{g}$ and $\mathbf{e}_{\mathrm{p}}$), antenna position error ($\mathbf{e}_{\mathrm{pos}}$), and mutual coupling error ($\mathbf{e}_{\mathrm{mc}})$. To this end, the definition of $\boldsymbol{a}\left(\theta_{k}, \boldsymbol{e}\right)$ would be as \cite{MLpaper} 
 \begin{equation}
 \begin{aligned}
  \boldsymbol{a}(\theta, \boldsymbol{e})&=\left(\boldsymbol{I}_{M}+\alpha_{\mathrm{mc}} \boldsymbol{E}_{\mathrm{mc}}\right) \times\left(\boldsymbol{I}_{M}+\operatorname{diag}\left(\alpha_{\mathrm{g}} \boldsymbol{e}_{\mathrm{g}}\right)\right) \\ \times & \operatorname{diag}\left(\exp \left(j \alpha_{\mathrm{e}} \boldsymbol{e}_{\mathrm{p}}\right)\right) \times \boldsymbol{a}\left(\theta, \alpha_{\mathrm{pos}} \boldsymbol{e}_{\mathrm{pos}}\right) \end{aligned}\label{steering_imperfections}
 \end{equation}
  \begin{align}
 \boldsymbol{e}_{\mathrm{g}}&=[0, \underbrace{0.2, \ldots, 0.2}_{\frac{M}{2}}, \underbrace{-0.2, \ldots,-0.2}_{\frac{M}{2}-1}]^{T} \label{e}
    \end{align}
 \begin{equation}
 \boldsymbol{e}_{\mathrm{p}}=\left[0, \underbrace{-\frac{\pi}{6}, \ldots,-\frac{\pi}{6}}_{\frac{M}{2}}, \underbrace{\frac{\pi}{6}, \ldots, \frac{\pi}{6}}_{\frac{M}{2}-1}\right]^{T} \label{ph}
 \end{equation}
 The position biases are
 \begin{equation}
 \boldsymbol{e}_{\mathrm{pos}}=[0, \underbrace{-0.2, \ldots,-0.2}_{\frac{M}{2}}, \underbrace{0.2, \ldots, 0.2}_{\frac{M}{2}-1}]^{T} \times d \label{pos}
 \end{equation}
 The mutual coupling coefficient vector is
 \begin{equation}
  \boldsymbol{e}_{\mathrm{mc}}=\left[0, \gamma^{1}, \ldots, \gamma^{19}\right]^{T} \label{mc}
 \end{equation} where $\gamma=0.3 e^{j \frac{\pi}{3}}$ is the mutual coupling coefficient between adjacent antennas, $\alpha_i \in [0,1], i \in \{g,p,\mathrm{pos},\mathrm{mc}\}$ is weighting parameter for each error. The choice of the error values in \eqref{e},\eqref{ph},\eqref{pos} and \eqref{mc} is system dependent, and can be changed accordingly.  $\boldsymbol{E}_{\mathrm{mc}}$ is defined as a Toeplitz matrix with parameter vector $ \boldsymbol{e}_{\mathrm{mc}}$ \cite{MLpaper}.
 In addition to array imperfections, in any real scenario, the received signals would be highly correlated, due to the contribution of multi-paths which makes $\boldsymbol{R}$ rank deficient or singular. Hence, the next section will define the model of coherent sources.
\subsection{Coherent Sources}
 In order to generate the multi-path component of each source, we consider having $K$ sources arriving from $K$ directions. At time instant n, there are $k$ transmit signals $x_k(t_n),$ $\forall k=1, \hdots ,K$, which arrive as replica of one of them i.e. $x_1(t_n)$, but phase delayed and magnitude weighted \cite{SSMUSIC}. Hence, the transmit waveform $x_{k}(t_n)$ in \eqref{errRx} can be redefined as
\begin{equation}
x_{k}(t_n)=g_{k} e^{j \phi_{k}} x_1(t_n), \quad k=1, \ldots, K,
\label{coherent}
\end{equation}
where $g_{k}$ is the amplitude factor of source $k$ and $\phi_{k}$ is the phase change of source $k$. Such model will impose rank deficiency on the covariance matrix structure in \eqref{covmatrix}, causing the existing DOA algorithms to fail accordingly.
To solve such problem, in the next section, we propose a deep auto-encoder to remove the effects of both coherence and array imperfections.

\section{DOA ESTIMATION BASED ON DEEP LEARNING}
\subsection{Deep Neural Network Architecture}
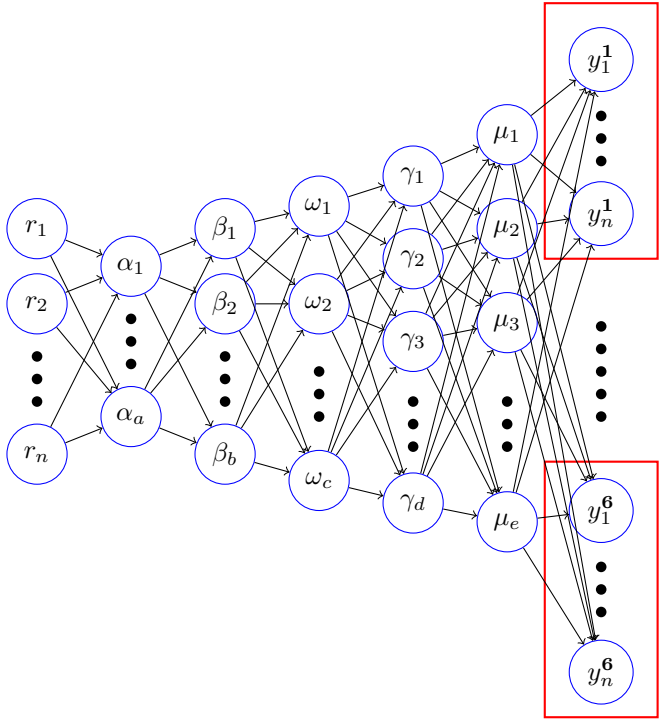
\begin{figure}[htbp]
	\begin{center}
		\begin{tikzpicture}
		\tikzstyle{place}=[circle, draw=blue, minimum size = 8mm]
		
		\foreach \x in {1,...,2}
		\node at (0, -1.5-\x*1) [place] (first_\x) {$r_\x$};
		\foreach \x in {1,...,3}
		\fill (0, -3.9 -\x*0.3) circle (2pt);
		\draw node at (0, -5*1.1) [place] (first_n) {$r_n$};
		
		\foreach \x in {1}
		\node at (1.25, -\x*3) [place] (second_\x){$\alpha_\x$};
		\foreach \x in {1,...,3}
		\fill (1.25, -3.4 -\x*0.3) circle (2pt);
		\draw node at (1.25, -5*1) [place] (second_m) {$\alpha_a$};
		
		\foreach \x in {1,...,2}
		\node at (2.5, -1.5-\x*1) [place] (third_\x){$\beta_\x$};
		\foreach \x in {1,...,3}
		\fill (2.5, -3.9 -\x*0.3) circle (2pt);
		\draw node at (2.5, -5*1.1) [place] (third_m) {$\beta_b$};
		
		\foreach \x in {1,...,2}
		\node at (3.75, -0.9-\x*1.3) [place] (forth_\x){$\omega_\x$};
		\foreach \x in {1,...,3}
		\fill (3.75, -4.1 -\x*0.3) circle (2pt);
		\draw node at (3.75, -5*1.17) [place] (forth_m) {$\omega_c$};
		
		\foreach \x in {1,...,3}
		\node at (5, -0.7-\x*1.1) [place] (fifth_\x){$\gamma_\x$};
		\foreach \x in {1,...,3}
		\fill (5, -4.5 -\x*0.3) circle (2pt);
		\draw node at (5, -5*1.23) [place] (fifth_m) {$\gamma_d$};
		
		\foreach \x in {1,...,3}
		\node at (6.25, -\x*1.25) [place] (sixth_\x){$\mu_\x$};

		\foreach \x in {1,...,3}
		\fill (6.25, -4.5 -\x*0.3) circle (2pt);

		\draw node at (6.25, -5*1.28) [place] (sixth_m) {$\mu_e$};
	
		\foreach \x in {1}
		\node at (7.5, 1-\x*1.25) [place] (seventh1_\x){$y^{\boldsymbol{1}}_{\x}$};
		\foreach \x in {1,...,3}
		\fill (7.5, -0.7-\x*0.3) circle (2pt);

		\node at (7.5, -2*1.15) [place] (seventh1_m) {$y^{\boldsymbol{1}}_{n}$};

		\foreach \x in {1,...,5}
		\fill (7.5, -3.5-\x*0.3) circle (2pt);
		\foreach \x in {1}
		\node at (7.5, -5-\x*1.25) [place] (seventh2_\x){$y^{\boldsymbol{6}}_{\x}$};
		\foreach \x in {1,...,3}
		\fill (7.5, -6.7-\x*0.3) circle (2pt);
		\node at (7.5, -6*1.4) [place] (seventh2_m) {$y^{\boldsymbol{6}}_{n}$};
		
		\node (rect) at (7.5,-7.3) [draw,thick,minimum width=1.5cm,minimum height=3.4cm,color=red] {};
		\node (rect) at (7.5,-1.2) [draw,thick,minimum width=1.5cm,minimum height=3.4cm,color=red] {};

		\foreach \i in {1,...,2}
		\foreach \j in {1}
		\draw [->] (first_\i) to (second_\j);
		\foreach \i in {1,...,2}
		\draw [->] (first_\i) to (second_m);
		\foreach \i in {1}
		\draw [->] (first_n) to (second_\i);
		\draw [->] (first_n) to (second_m);
		
		\foreach \i in {1}
		\foreach \j in {1,...,2}
		\draw [->] (second_\i) to (third_\j);
		\foreach \i in {1}
		\draw [->] (second_\i) to (third_m);
		\foreach \i in {1,...,2}
		\draw [->] (second_m) to (third_\i);
		\draw [->] (second_m) to (third_m);
		
		\foreach \i in {1,...,2}
		\foreach \j in {1,...,2}
		\draw [->] (third_\i) to (forth_\j);
		\foreach \i in {1,...,2}
		\draw [->] (third_\i) to (forth_m);
		\foreach \i in {1,...,2}
		\draw [->] (third_m) to (forth_\i);
		\draw [->] (third_m) to (forth_m);
		
		\foreach \i in {1,...,2}
		\foreach \j in {1,...,3}
		\draw [->] (forth_\i) to (fifth_\j);
		\foreach \i in {1,...,2}
		\draw [->] (forth_\i) to (fifth_m);
		\foreach \i in {1,...,3}
		\draw [->] (forth_m) to (fifth_\i);
		\draw [->] (forth_m) to (fifth_m);
		
		\foreach \i in {1,...,3}
		\foreach \j in {1,...,3}
		\draw [->] (fifth_\i) to (sixth_\j);
		\foreach \i in {1,...,3}
		\draw [->] (fifth_\i) to (sixth_m);
		\foreach \i in {1,...,3}
		\draw [->] (fifth_m) to (sixth_\i);
		\draw [->] (fifth_m) to (sixth_m);
		
		\foreach \i in {1,...,3}
		\foreach \j in {1}
		\draw [->] (sixth_\i) to (seventh1_\j);
		\foreach \i in {1,...,3}
		\draw [->] (sixth_\i) to (seventh1_m);
		\draw [->] (sixth_m) to (seventh1_m);
		\draw [->] (sixth_m) to (seventh1_1);
		
		\foreach \i in {1,...,3}
		\foreach \j in {1}
		\draw [->] (sixth_\i) to (seventh2_\j);
		\foreach \i in {1,...,3}
		\draw [->] (sixth_\i) to (seventh2_m);
		\draw [->] (sixth_m) to (seventh2_m);
		\draw [->] (sixth_m) to (seventh2_1);
		
%
		\end{tikzpicture}
		\vspace{10pt}
		\caption{Proposed autoencoder consisting of 5 hidden layers with hidden layer 1 acts like an encoder and the following layers combined act like a decoder}
		\label{autoencoder_architecture}
	\end{center}
\end{figure}

 A deep neural network (DNN) is an artificial neural network (ANN) with multiple layers between the input and output layers. The DNN learns the correct mathematical manipulation to map the input into the output. It has the ability to solve complex nonlinear problems \cite{b4}.
Here, we propose a DNN architecture which is an autoencoder (AE), where the first hidden layer performs the function of an encoder as it reduces the dimension of the input by extracting the main features of the input. Afterwards, the encoding layer is followed by four hidden layers that help in the decoding process by retrieving the information to restore back the input of the AE. Table \ref{table1} shows the size of all the hidden layers. The output layer consists of six sub-regions where each region is considered as decoder by itself. Therefore, the  AE has six decoders and each decoder retrieves specific information from the input, as shown in Fig. \ref{autoencoder_architecture}. The process of encoding-decoding helps decrease the impact of disturbances in the autoencoder input, through de-noising the input, retrieving only the useful information. The disturbance is our case is mainly due to array imperfections, noise, and coherent sources.

\begin{table}[htbp]
       \begin{center}
       \begin{tabular}{ |c|c| } 
       \hline
       Layer & Size \\
       \hline
       Input Layer & $n=$ 380 \\
       \hline
       Hidden Layer 1 & $a=$ 190 \\
       \hline
       Hidden Layer 2 & $b=$ 380 \\
       \hline
       Hidden Layer 3 & $c=$ 570 \\
       \hline
       Hidden Layer 4 & $d=$ 760 \\
       \hline
       Hidden Layer 5 & $e=$ 950 \\
       \hline
       Output Layer & $6\times380 = 2280$ \\
       \hline
       \end{tabular}
       \end{center}
          \caption{Sizes of all layers}
       \label{table1}
\end{table}
\subsection{Learning Scheme}

The input of the proposed DNN is the covariance matrix of the received signal as defined by \eqref{covmatrix}.
In order to reduce the dimension of the input layer, we consider only the correlated elements in the covariance matrix. By exploiting the symmetry in the correlation matrix, only the strict upper or strict lower triangular part of the matrix can be considered \cite{MLpaper}. In this paper, the off-diagonal upper right matrix of the covariance matrix is considered. For example, if we have $M=3$ antennas, the covariance matrix will be

\begin{equation}
\mathbf{R}=\left[\begin{array}{lll}{r_{11}} & {r_{12}} & {r_{13}} \\ {r_{21}} & {r_{22}} & {r_{23}} \\ {r_{31}} & {r_{32}} & {r_{33}}\end{array}\right],
\end{equation}
then the following input is obtained 
\begin{equation}
{\hat{\boldsymbol{r}}}=\left[\begin{array}{lll}{r_{12}} & {r_{13}} & {r_{23}} \end{array}\right]^{T}.
\end{equation}
Generally, let $\mathbf{R} \in \mathbb{C}^{M \times M}$, then the input vector is ${\hat{\boldsymbol{r}}} \in  \mathbb{C}^{M(M-1) / 2}$. Additionally, the input of the DNN must be real valued, hence, ${\hat{\boldsymbol{r}}}$ is converted from complex to real, by concatenating $\operatorname{Real}\left\{\hat{\boldsymbol{r}}^{T}\right\}$ with $\operatorname{Imag}\left\{\hat{\boldsymbol{r}}^{T}\right\}$ to produce $\mathbf{r} \in  \mathbb{C}^{n}$, where $n= M(M-1)$.
\begin{equation}
\boldsymbol{r}=\left[\operatorname{Real}\left\{\hat{\boldsymbol{r}}^{T}\right\}, \operatorname{Imag}\left\{\hat{\boldsymbol{r}}^{T}\right\}\right]^{T}.
\end{equation}
The proposed autoencoder in Fig.\ref{autoencoder_architecture} decomposes its input into $6$ spatial subregions, each spatial subregion is a specific range of angles and all subregions are of the same size. To define these subregions, $7$ particular spatial angles were chosen such as
    
\begin{equation}
\theta^{(1)}<\theta^{(1)}<\cdots<\theta^{(7)}
\end{equation}
with constant gaps such that $\theta^{i+1}-\theta^{i}=\mathrm{constant}$ $\forall i, i=1, \hdots,7$,
and each subregion $j$ is defined as $\left[\theta^{\mathrm{j}},\theta^{\mathrm{j+1}}\right]$ where $j = 1, \ldots, 6$.
Therefore, if the input vector of the autoencoder $\boldsymbol{r}_k$ is generated using a signal impinging from source $k$  on the antenna array at angle $\Theta_{k}$ within the $j$-th subregion, then the output of the $j$-th decoder will be $\boldsymbol{r}_k$, while the output of the other decoders will be zero as there are no signals impinging from those range of angles.
The AE is trained to be able to separate multiple signals transmitted from sources located in different subregions impinging onto the array simultaneously. Hence, it is able to decompose the input vector with components belonging to different subregions and extract the information that belongs to every subregion by retrieving it in the related decoders. 
\subsection{Training Process}
The data set is constructed by generating $I$ training samples of the covariance vector $\boldsymbol{r}$ corresponding to single-signal scenarios. The data was generated with random angles that spans all subregions. The output of each decoder is determined based on which subregion the generated $\boldsymbol{r}$ belong to. Alternatively, the decoder can be considered as a spatial filter. This filter extracts the covariance vector information that belongs to a particular subregion. 
In order to build up the training label of the entire output of the AE, the outputs of the six decoders are concatenated as follows

\begin{align}
\boldsymbol {y}=& [\boldsymbol {y}_{1}^{T}, \ldots, \boldsymbol {y}_{6}^{T}]^{T} \nonumber \\=&\left [{{\boldsymbol {0}}, \ldots, {\boldsymbol {0}} , \underbrace{\boldsymbol {r}^{T}(\Theta _{k})}_{j\text{ th subregion}}, {\boldsymbol {0}}, \ldots, {\boldsymbol {0}}} \right]^{T},
\label{training_label}
\end{align}
in which $\boldsymbol {y}_j$ is the output of the $j$ decoder.
To train the autoencoder, the squared $l_2$-norm distance between the actual output and the expected one is used as the loss function. That is,

\begin{equation*} \boldsymbol {L }(\Theta _{k}) = \frac {1}{2} \|\tilde {\boldsymbol {y}}(\Theta _{k}) \|_{2} ^ {2}
\end{equation*}        
where
\begin{equation*} \tilde {\boldsymbol {y}}(\Theta _{k}) = \boldsymbol {y}(\Theta _{k}) -\hat {\boldsymbol {y}}(\Theta _{k})
\end{equation*}
and $\hat{\boldsymbol{y}}\left(\Theta_{k}\right)$ is the actual output of the autoencoder when $\boldsymbol{r}\left(\boldsymbol{\Theta}_{k}\right)$ is the input. The optimizer used in the training process to minimize the loss function is a RMSProp optimizer \cite{MLpaper}.

\subsection{Scanning}
After training the network shown in Fig. \ref{autoencoder_architecture}, the training label in \eqref{training_label} is used to estimate the original directions of the correlated sources impinging on the array. This is done through spatial scanning of the output of each decoder (i.e. filter) $\boldsymbol {y}_j$. The scanning phase aims at calculating the gain of each filter in all directions, in which the actual source angles would have large gain values, while the other directions would have much smaller gains. Afterwards, a threshold value is used to select the angles whose gains have peaks surpassing the threshold. The gain response of each filter is obtained by
\begin{equation}
g^{(j)}=\left|\overline{\boldsymbol{r}}^{H}\left(\Theta_{k}\right) \overline{\boldsymbol{y}}_{j}\right|, \quad  j=1, \ldots, 6,
\label{gain}
\end{equation}
where the superscript $(\bullet)^{H}$ is the conjugate transpose of the matrices and vectors, $\overline{\boldsymbol{r}}^{H}$ is the estimated value of ${\hat{\boldsymbol{r}}}$, and  $\overline{\boldsymbol{y}}_{j}$ is the complex version of the output of the $j$-th decoder $\boldsymbol{y}_{j}$. $\overline{\boldsymbol{y}}_{j}$ is obtained from $\boldsymbol{y}_{j}$ by concatenating the first half that represents the real values in $\boldsymbol{y}_{j}$ with their corresponding imaginary values in the second half of $\boldsymbol{y}_{j}$, similarly $\overline{\boldsymbol{r}}^{H}\left(\Theta_{k}\right)$ is obtained from $\boldsymbol {r}^{T}(\Theta _{k})$ .

\section{Simulation Results}
In this section, we carry out simulations to evaluate our proposed DNN. We used the python library \textit{tensorflow} to design and process our DNN. The network is trained on $I=1200$ samples, with learning rate of $0.001$, while the batch size is $100$ and the number of epochs is $1000$ epochs. We use a uniform linear array (ULA) of size $M=20$ elements with spacing $d=\lambda/2$ to predict directions of signals impinging from sources located in the spatial range of $\left[-60^{\circ}, 60^{\circ}\right]$, which is divided equally into six subregions. The training samples are generated randomly from directions $\Theta_{i}=-60^{\circ}+0.1^{\circ}, \forall i=1, i=1,\hdots,1200$. The covariance input vector $\boldsymbol{r}\left(\Theta_{i}\right) $ is generated using $N=800$ snapshots.  To evaluate the performance of our algorithm, we use forward/backward spatial smoothing along with the MUSIC algorithm (SS-MUSIC) in \cite{SSMUSIC2}, and compared it against our DNN in multipath environment by randomly changing $g_k$ and $\phi_{k}$ in \eqref{coherent} for every target $k$. 

\subsection{Gain responses of each decoder (i.e filter)}
\InputIfFileExists{response10.tikz}{}

The DNN is tested using a covariance vector obtained from two correlated sources located at $\theta_{1}=-15^{\circ}, \theta_{2}=-5^{\circ}$ respectively, which belongs to subregion 3, i.e. $\left[\mathrm{-20}^{\circ},\mathrm{0}^{\circ}\right]$, with signal to noise ratio $\mathrm{SNR}= 10$ dB. Those specific directions were chosen, because they are in the same subregion, making them spatially close, hence it would be harder to separate them compared to distant sources. The threshold value to find the peaks in the scanning process is set to $= 0.3$, which is set by experience. Fig. \ref{gain response} shows the gain obtained from \eqref{gain} for all the filters. It can be depicted that the spatial gain response of the filter corresponding to the assigned subregion has higher peaks compared to the other filters. It is clear that the filter managed to differentiate between both angles despite the fact they are coherent, and the antenna array suffers from imperfection errors as stated in \eqref{steering_imperfections}. 
\subsection{Performance against SS-MUSIC }
\InputIfFileExists{mse.tikz}{}

Fig.\ref{mse} compares DOA estimation performance of our proposed DNN with SS-MUSIC in presence of correlated sources and array imperfections combined. The average root mean square error (RMSE) in degrees is used to measure the accuracy of DOA estimates for various $\mathrm{SNR}$. It can be shown from the figure that the RMSE of our DNN starts high at $\mathrm{SNR} = 0$ dB, then decreases significantly compared to SS-Music as the $\mathrm{SNR}$ increases.
It can be seen from the figure that the RMSE of SS-MUSIC algorithm is heavily impacted by the presence of imperfections due to the fact that it assumes ideal steering vector model with no imperfections as in \eqref{received_signal}
\subsection{Perfomance against algorithm in \cite{MLpaper} }
\InputIfFileExists{compare.tikz}{}
\\
Fig. \ref{compare} compares the detection performance of our DNN compared to the algorithm in \cite{MLpaper}, where the authors only considered array imperfections and assumes perfectly uncorrelated sources for DOA estimation. The same training data was used for both algorithms for fair comparison. The figure shows a consistent behavior for our algorithm detecting all 8 targets, however the algorithm in \cite{MLpaper} shows lower probability of detection due to its failure to detect all coherent sources.
\section{Conclusion}
In this paper we presented a framwork for DOA estimation of correlated sources in presence of array imperfections. Our approach was based on Deep AE with 5 hidden layers, one acting as encoder and 4 as decoder. From the simulations we showed that the AE acts as a denoiser, where it could successfully remove the effect of coherence and imperfections producing accurate DOA estimates compared to the commonly used SS-MUSIC. Moreover, we compared our algorithm with the approach in \cite{MLpaper} where the authors only deal with imperfections, and our algorithm showed better and more consistent behavior.
\section*{Acknowledgment}
Funded by the Deutsche Forschungsgemeinschaft (DFG, German Research Foundation) – Project-ID 287022738 – TRR 196 S03
%
%
%
%
%
%
\bibliographystyle{IEEEtran}
\bibliography{conf}

\begin{thebibliography}{10}
\providecommand{\url}[1]{#1}
\csname url@samestyle\endcsname
\providecommand{\newblock}{\relax}
\providecommand{\bibinfo}[2]{#2}
\providecommand{\BIBentrySTDinterwordspacing}{\spaceskip=0pt\relax}
\providecommand{\BIBentryALTinterwordstretchfactor}{4}
\providecommand{\BIBentryALTinterwordspacing}{\spaceskip=\fontdimen2\font plus
\BIBentryALTinterwordstretchfactor\fontdimen3\font minus
  \fontdimen4\font\relax}
\providecommand{\BIBforeignlanguage}[2]{{%
\expandafter\ifx\csname l@#1\endcsname\relax
\typeout{** WARNING: IEEEtran.bst: No hyphenation pattern has been}%
\typeout{** loaded for the language `#1'. Using the pattern for}%
\typeout{** the default language instead.}%
\else
\language=\csname l@#1\endcsname
\fi
#2}}
\providecommand{\BIBdecl}{\relax}
\BIBdecl

\bibitem{mvdr}
J.~Li, Z.~Wang, and P.~Stoica, \emph{Robust Adaptive Beamforming}, 2005, ch.~3,
  pp. 91--200.

\bibitem{b7}
S.~Theodoridis and R.~Chellappa, \emph{Academic Press Library in Signal
  Processing, Volume 3: Array and Statistical Signal Processing}, 1st~ed.\hskip
  1em plus 0.5em minus 0.4em\relax Orlando, FL, USA: Academic Press, Inc.,
  2013.

\bibitem{b3}
P.~Stoica, J.~Li, and H.~He, ``Spectral analysis of nouniformly sampled data: A
  new approach versus the periodogram,'' \emph{Trans. Sig. Proc.}, vol.~57,
  no.~3, p. 843–858, Mar. 2009.

\bibitem{SSMUSIC}
{Qing Chen} and {Ruolun Liu}, ``On the explanation of spatial smoothing in
  music algorithm for coherent sources,'' in \emph{International Conference on
  Information Science and Technology}, March 2011, pp. 699--702.

\bibitem{fried}
B.~{Friedlander} and A.~J. {Weiss}, ``Direction finding in the presence of
  mutual coupling,'' \emph{IEEE Transactions on Antennas and Propagation},
  vol.~39, no.~3, pp. 273--284, March 1991.

\bibitem{MCsol1}
T.~{Svantesson}, ``Modeling and estimation of mutual coupling in a uniform
  linear array of dipoles,'' in \emph{1999 IEEE International Conference on
  Acoustics, Speech, and Signal Processing. Proceedings. ICASSP99}, vol.~5,
  March 1999, pp. 2961--2964.

\bibitem{MCsol2}
M.~{Lin} and L.~{Yang}, ``Blind calibration and {DOA} estimation with uniform
  circular arrays in the presence of mutual coupling,'' \emph{IEEE Antennas and
  Wireless Propagation Letters}, vol.~5, pp. 315--318, 2006.

\bibitem{MLpaper}
Z.~{Liu}, C.~{Zhang}, and P.~S. {Yu}, ``Direction-of-arrival estimation based
  on deep neural networks with robustness to array imperfections,'' \emph{IEEE
  Transactions on Antennas and Propagation}, vol.~66, no.~12, pp. 7315--7327,
  Dec 2018.

\bibitem{b4}
J.~Schmidhuber, ``Deep learning in neural networks: An overview,'' \emph{Neural
  Networks}, vol.~61, pp. 85 -- 117, 2015.

\bibitem{SSMUSIC2}
U.~Pillai and B.~Kwon, ``\BIBforeignlanguage{English (US)}{Forward/backward
  spatial smoothing techniques for coherent signal identification},''
  \emph{\BIBforeignlanguage{English (US)}{IEEE Transactions on Signal
  Processing}}, vol.~37, no.~1, pp. 8--15, 1 1989.

\end{thebibliography}

\end{document}